\documentclass{article}
\usepackage{calor2kpro}
\usepackage{epsfig}
\begin{document}
\title{ 
  PRECISE MEASUREMENT OF JET ENERGIES WITH THE ZEUS DETECTOR
  }
\author{
  Matthew Wing \\
  (On behalf of the ZEUS Collaboration)\\	
  {\em McGill University,}\\
  {\em Physics Department,} \\
  {\em 3600 University Street,} \\
  {\em Montreal, Canada, H3A 2T8}
  }
\maketitle
\baselineskip=11.6pt
\begin{abstract}
  A method to correct the jet transverse energy has been developed for the ZEUS detector which 
  attains an uncertainty better than 3$\%$. The procedure is based on a combination of tracking 
  and calorimeter information that optimises the resolution of reconstructed kinematic variables. 
  The selected calorimeter clusters and tracks are referred to as Energy Flow Objects (EFOs). 
  The conservation of energy and momentum in neutral current deep inelastic $e^+p$ scattering 
  events is exploited to determine the required energy corrections by balancing the scattered 
  positron with the hadronic final state. The method has been applied to data and simulated 
  events independently. The corrected EFOs are used as input to a $k_T$-cluster jet algorithm 
  to reconstruct the jets and to determine kinematic variables. In addition, the corrected EFOs
  allow an improved measurement of the internal structure of a jet.
\end{abstract}
\baselineskip=14pt
%
%
\section{Introduction}

Through studying jet production at the $ep$ collider, HERA, a wide variety of fundamental 
measurements are possible, such as the determination of the strong coupling constant, 
$\alpha_s$, or information on the gluon content of the photon. To make the measurements as 
significant as possible, jets have to be reconstructed, optimising the energy resolution and 
minimising the uncertainty of the absolute energy scale. The accurate reconstruction of jets 
at the ZEUS experiment generally relies on the precise determination of energies and angles 
of hits in the uranium-scintillator calorimeter (CAL)\cite{CAL}. Before reaching the CAL, 
particles pass through approximately one radiation length of dead material in the central 
region of the detector mainly due to the solenoid magnet between the tracking detector and 
the CAL. The modelling of the energy lost by particles traversing the solenoid plays the most 
significant r\^{o}le in understanding the absolute energy scale between data and MC.

It has been shown\cite{zeus_nc} that the energy scale of high-energy scattered positrons 
in the central part of the CAL is understood to within 1$\%$. Similarly, the hadronic energy 
scale (i.e. the energy carried by particles in the neutral current (NC) final state, excluding 
the scattered positron) is known to within 2$\%$, of which 1$\%$ comes from the uncertainty 
of the positron measurement. The uncertainty when specifically studying jet production (jets 
in the range of transverse energy, $E_T^{\rm jet} \sim 10-100$ GeV) is not the same 
as the hadronic energy scale because the choice of algorithm and modelling of the jet structure 
can affect the result. The absolute energy scale uncertainty for jet production is currently 
known to within 3$\%$\cite{dijet}, which contributes a roughly 15$\%$ error on a jet cross 
section measurement, due to the steeply falling $E_T^{\rm jet}$ distribution. 

This article describes a method capable of reducing the jet energy scale uncertainty to 
$1-2\%$ as for the case of the electromagnetic and hadronic energy scale uncertainties. 
First, the reconstruction of Energy Flow Objects (EFOs) is introduced, which optimise the use 
of tracking and CAL information. Then, a method is outlined which exploits energy and 
momentum conservation in a high purity sample of NC events to correct the CAL-EFOs for 
energy loss in the dead material of the detector.

%
%
\section{Jet energy correction method}

\subsection{Energy flow objects}

The use of EFOs has been shown to improve the reconstruction of kinematic 
quantities\cite{briskin}. Clusters of cells are formed and combined with tracks originating 
from the primary vertex. In general, for low momentum particles, tracks provide a better 
measurement of the energy, whereas the CAL is better for high momentum particles. In the case 
of clusters not matched to a track (from, for example, neutral particles) or tracks not matched 
to a cluster (where the particle momentum is so small that it does not reach the CAL), the 
situation is clear. However, for clusters and tracks considered matched (or even many tracks 
matched to one cluster, etc.), a decision has to be made concerning which quantity (or 
quantities) to use. For a matched cluster-track system, the resolutions and ratio of energy 
to momentum are considered in the decision-making process. Using this procedure, a list of 
track-EFOs and CAL-EFOs is obtained. The track-EFOs are assumed to be an accurate measurement 
of the particle energy, whereas the CAL-EFOs are subject to energy-loss in the dead material 
in front of the CAL and must be corrected.

\subsection{Sample of neutral current events}

The conservation of energy and momentum in NC events can be exploited to determine the 
CAL-EFO energy-correction functions by balancing the momentum of the scattered positron with 
that of the hadronic final state. Using a method similar to that detailed below, the 
uncertainty in the ZEUS jet energy-scale has already been determined to within 
3$\%$\cite{joost}. In order to reduce this uncertainty, the form of the function for energy 
loss, improving the samples of events chosen and using a larger data sample have been 
studied\cite{andreas}.

Two samples of events with high momentum transfer, $Q^2>100$ GeV$^2$, were used to provide 
full angular coverage of the detector. A sample of events with high positron $p_T$ and a sample 
of events at high effective longitudinal momentum, $y$, were used. The kinematic variables of 
the positron were reconstructed using the double angle method\cite{DA}, which, to first order, 
is independent of the absolute energy scale of the CAL. The hadronic final state four-vector 
was calculated from the EFOs reconstructed as above and its momentum components balanced with 
that of the scattered positron. Each correction for the CAL-EFOs were parametrised as a 
function of energy. There parameters were determined for several bins of polar angle, $\theta$,  
(reflecting the detector geometry), by minimisation of the following quantity:

\[ {\sum_{\rm sample \ 1} min \left[ \left( \frac{p_T^{\rm DA} - p_T^{\rm had}}
                                                {p_T^{\rm DA}} \right)^2, 0.2^2 \right]
 + \sum_{\rm sample \ 2} min \left[ \left( \frac{y^{\rm DA} - y^{\rm had}}
                                                {y^{\rm DA}} \right)^2, 0.2^2 \right]}. \]

The minimisation was performed using the MINUIT package\cite{minuit} and correction factors 
obtained separately for data and MC. The difference between data and MC may arise from 
inadequate detail in the description of the dead material in front of the CAL and an 
inaccurate simulation of the hadronic energy loss process.

\subsection{Energy correction functions}

\begin{figure}[ht]
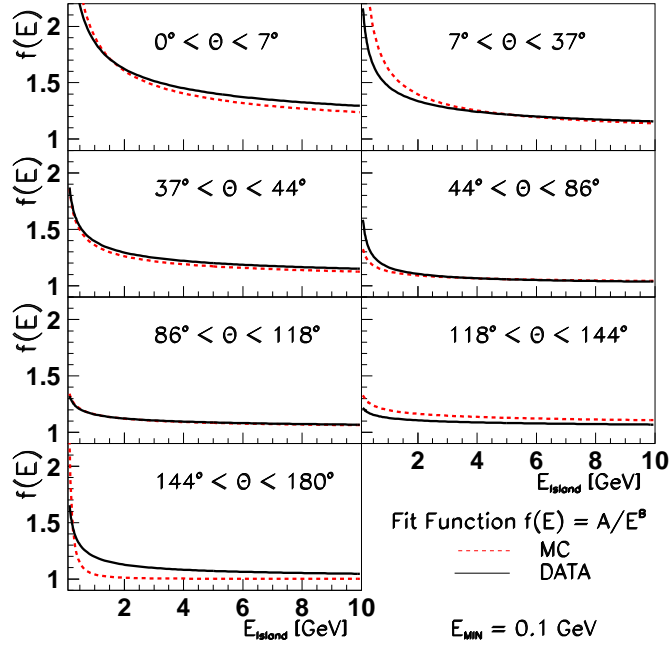

\vspace{-0.75cm}
\caption{Energy corrections as a function of cluster energy in bins of $\theta$. 
The corrections are shown separately for data (solid line) and MC (dashed line).}
\label{fitfunc}
\end{figure}

The CAL clusters were corrected for energy loss using the functional form, $f(E)~=~A/E^B$, 
where $E$ is the cluster energy and $A$ and $B$ are the correction factors to be determined in 
each angular region. The result of the fits for both data and {\sc Herwig} MC\cite{herwig} are 
shown in fig.~\ref{fitfunc}. It can be seen that the data and MC show similar trends but differ 
in detail, justifying the need to perform the fits and apply the corrections separately for data 
and MC.

%
%
\section{Results}

\begin{figure}[ht]
\unitlength=1mm
\begin{picture}(0,0)(100,100)
\put(170,87){\bf \Large{(a)}}
\put(170,46){\bf \Large{(b)}}
\end{picture}
\begin{center}
\centerline{~\epsfig{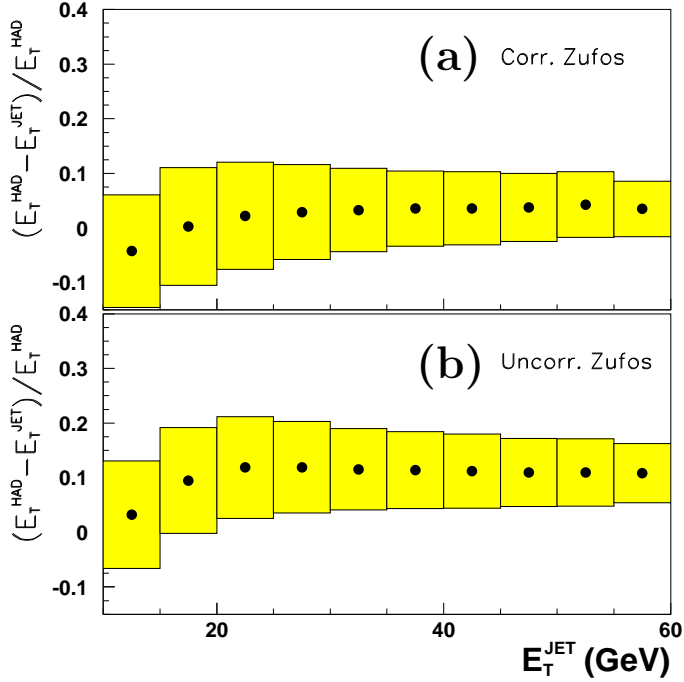}}
\end{center}
\vspace{-0.85cm}
\caption{Fractional difference between hadron-level jet $E_T$ and that 
reconstructed with (a) corrected EFOs and (b) uncorrected EFOs as a function of the 
transverse energy. The shaded band shows the width of the distribution.}
\label{ethad}
\end{figure}

To test the validity of the factors obtained, the correction functions were applied to an 
independent photoproduction MC sample, where the scattered positron is not detected in the CAL. 
Jet quantities were reconstructed using both EFOs with and without correction and the 
transverse energy, $E_T^{\rm JET}$, compared to the ``true'' hadron-level, $E_T^{\rm HAD}$ 
as shown in fig.~\ref{ethad}.   

Before the EFOs are corrected, the deviation from the true value is roughly $10-15\%$ as 
shown in fig.~\ref{ethad}b. After correction, as in fig.~\ref{ethad}a, the transverse 
energies are significantly closer to the true values, demonstrating that the energy 
correction helps to reproduce the true quantities when applied to an independent MC 
sample. The extent to which hadron-level quantities in the data are reproduced and the size 
of the resulting energy scale uncertainty can be determined by comparing the correction in 
data and MC separately.

\begin{center}
~\epsfig{file=fitfunc.epsi,height=8.5cm}
\end{center}

\begin{figure}[ht]
\unitlength=1mm
\begin{picture}(0,0)(100,100)
\put(196,88){\bf \Large{(a)}}
\put(195,47){\bf \Large{(b)}}
\end{picture}
\begin{center}
~\epsfig{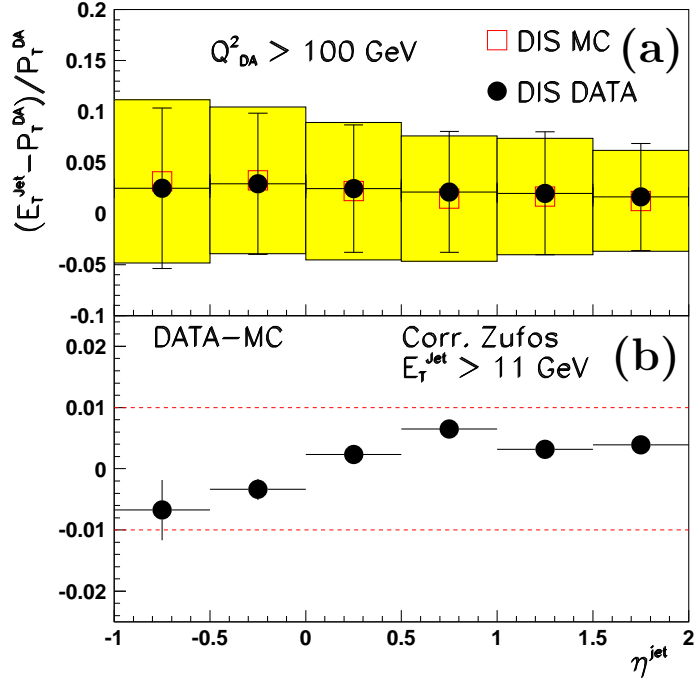}
\end{center}
\vspace{-0.75cm}
\caption{(a) Fractional difference of $E_T^{\rm Jet}$ and $p_T^{\rm DA}$ for data and MC and 
(b) the difference, DATA$-$MC.}
\label{etjet}
\end{figure}

The results of applying the corrections to the same data and MC used to obtain the factors 
is shown in fig.~\ref{etjet}. Here, instead of the $p_T$ of the hadronic final state, the 
transverse energy of the jet, $E_T^{\rm Jet}$, which takes into account the extra 
uncertainties due to jet reconstruction and is closely related to $p_T^{\rm HAD}$, is 
considered. As a function of the pseudorapidity of the jet, $\eta^{\rm Jet}$ 
($= -\ln[\tan(\theta/2)]$), the data and MC agree to within roughly $1\%$.

For the global kinematic quantities, $p_T$ and $y$, a similar trend was seen, i.e. 
the data and MC agreed to within $1-2\%$ in all parts of the kinematic region considered.

%
%
\section{Conclusions}

A method has been developed for correcting for energy lost in the dead material in front of the 
ZEUS calorimeter. The procedure relies on a combination of tracking and CAL information. The 
CAL energy correction is determined using energy and momentum balanced NC events. 
A more accurate reproduction of the hadronic final state is obtained indicating that 
the absolute CAL energy scale is reproduced to within $1-2\%$ between data and MC.

\end{document}